\documentclass[ reprint,amsmath,amssymb, aps,prl]{revtex4-2} 

\usepackage{amsmath}
\usepackage{amssymb}
\usepackage{lipsum}
\usepackage[cjk]{kotex}
\usepackage{graphicx}
\usepackage{dcolumn}
\usepackage{xcolor}
\usepackage{bm}
\usepackage[mathscr]{euscript}
\usepackage{comment}
\usepackage{tikz}
\usetikzlibrary{quotes,angles}

\newcommand{\VBone}{\kern.08em
\begin{tikzpicture}[baseline={([yshift=-0.5ex]current bounding box.center)}]
    \draw coordinate (a) at (0,0);
    \draw coordinate (b) at (0.4,0);
    \draw coordinate (c) at (0.2,0.35);
    \draw (a) -- (b) -- (c) -- (a) pic [draw=black]{} ;
    \draw[color=black!100] (c) circle (0.04);
    \fill[magenta!100] (0.2,0) ellipse (0.23 and 0.06);
\end{tikzpicture}
\kern.08em%
}

\newcommand{\VBtwo}{\kern.08em
\begin{tikzpicture}[baseline={([yshift=-0.5ex]current bounding box.center)}]
    \draw coordinate (a) at (0,0);
    \draw coordinate (b) at (0.4,0);
    \draw coordinate (c) at (0.2,0.35);
    \draw (a) -- (b) -- (c) -- (a) pic [draw=black]{} ;
    \draw[color=black!100] (a) circle (0.04);
    \fill[magenta!100] (0.3,0.17) ellipse [x radius = 0.23, y radius =  0.06, rotate = 120];
\end{tikzpicture}
\kern.08em%
}

\newcommand{\VBthree}{\kern.08em
\begin{tikzpicture}[baseline={([yshift=-0.5ex]current bounding box.center)}]
    \draw coordinate (a) at (0,0);
    \draw coordinate (b) at (0.4,0);
    \draw coordinate (c) at (0.2,0.35);
    \draw (a) -- (b) -- (c) -- (a) pic [draw=black]{} ;
    \draw[color=black!100] (b) circle (0.04);
    \fill[magenta!100] (0.1,0.17) ellipse [x radius = 0.23, y radius =  0.06, rotate = 60];
\end{tikzpicture}
\kern.08em%
}

\newcommand{\DimerOne}{\kern.08em
\begin{tikzpicture}[baseline={([yshift=-0.5ex]current bounding box.center)}]
    \draw coordinate (a) at (0,0);
    \draw coordinate (b) at (0.4,0);
    \draw coordinate (c) at (0.2,0.35);
    \draw (a) -- (b) -- (c) -- (a) pic [draw=black]{} ;
    \draw[color=black!100] (c) circle (0.04);
    \fill[black!100] (0.2,0) ellipse (0.23 and 0.06);
\end{tikzpicture}
\kern.08em%
}

\newcommand{\DimerTwo}{\kern.08em
\begin{tikzpicture}[baseline={([yshift=-0.5ex]current bounding box.center)}]
    \draw coordinate (a) at (0,0);
    \draw coordinate (b) at (0.4,0);
    \draw coordinate (c) at (0.2,0.35);
    \draw (a) -- (b) -- (c) -- (a) pic [draw=black]{} ;
    \draw[color=black!100] (a) circle (0.04);
    \fill[black!100] (0.3,0.17) ellipse [x radius = 0.23, y radius =  0.06, rotate = 120];
\end{tikzpicture}
\kern.08em%
}

\newcommand{\DimerThree}{\kern.08em
\begin{tikzpicture}[baseline={([yshift=-0.5ex]current bounding box.center)}]
    \draw coordinate (a) at (0,0);
    \draw coordinate (b) at (0.4,0);
    \draw coordinate (c) at (0.2,0.35);
    \draw (a) -- (b) -- (c) -- (a) pic [draw=black]{} ;
    \draw[color=black!100] (b) circle (0.04);
    \fill[black!100] (0.1,0.17) ellipse [x radius = 0.23, y radius =  0.06, rotate = 60];
\end{tikzpicture}
\kern.08em%
}

\begin{document}
\preprint{APS/123-QED}

\title{Exact Hole-induced 
Resonating-Valence-Bond Ground State in Certain $U=\infty$ Hubbard Models}

\author{Kyung-Su Kim}
 \altaffiliation[]{kyungsu@stanford.edu}
\affiliation{Department of Physics, Stanford University, Stanford, CA 93405}%

\date{\today}

\begin{abstract}
We prove that the motion of a single hole induces the nearest-neighbor resonating-valence-bond (RVB) ground state in the $U=\infty$ Hubbard model on
a triangular cactus -- a tree-like variant of a kagome lattice.
The result can be easily generalized to $t-J$ models with antiferromagnetic interactions $J\geq 0$ on the same graphs.
This is a {weak} converse of
Nagaoka's theorem of  ferromagnetism on a bipartite lattice. \end{abstract}

\maketitle

  A resonating-valence-bond (RVB) state is an exotic spin liquid state originally envisioned by Anderson \cite{
anderson1973RVB}.
It was revisited after the discovery of high $T_c$ superconductivity 
\cite{anderson1987RVB, baskaran1993RVB}, which gave rise to the notion that by doping the RVB, holons, the fractionalized excitations carrying charge $e$ and spin $0$, can condense to become a superconductor \cite{rokhsarKivelson1988QDM, kivelsonRokhsarSethna, kivelsonRokhsar}.
In this picture, the background antiferromagnetic interaction, $J$, plays an essential role  as a mediator of valence-bond formation and thus of ``preformed Cooper pairs."

Even in the absence of explicit
exchange interactions, however, 
magnetism can still arise upon doping of the Hubbard model at half-filling in the $U=\infty$ limit (where $J=0$).
The idea is that the motion of a doped hole (or electron) shuffles the background spin ordering, leading to the magnetism \cite{thouless1965exchange}.
In particular, the celebrated ``Nagaoka's theorem" states that for a bipartite system (e.g. a square lattice), introducing a single hole leads to a fully polarized ferromagnetic ground state due to the constructive interference of the hole motion in a ferromagnetic background \cite{nagaoka1966ferromagnetism}.
This result was generalized to a wider class of graphs by Tasaki \cite{tasaki1989extension} -- the only requirement is that the product of hopping matrix elements around any loop in the graph is positive. (See also \cite{kim2022interstitial, moessner2000slowHole} for a related theme on kinetically induced magnetism.)
On a non-bipartite lattice, however, the product of hopping matrix elements around loops with an odd number of bonds is negative, frustrating the kinetic energy of a hole in a ferromangetic background.
Indeed, recent numerical studies have concluded that the ground state of the $U=\infty $ Hubbard model on a triangular lattice in the presence of a single hole has total spin zero ($S_{\rm tot} =0$) and has 120$^{\circ}$ order as in the case of triangular lattice antiferromagnet \cite{haerter2005kineticAF, sposetti2014kineticAF, lisandrini2017kineticAF, zhu2022doped}.

In this paper, starting from a simple problem on a single triangle, we study the $U=\infty$ Hubbard model on a certain class of graphs known as a triangular cactus (also known as a Husimi cactus), on which the kinetic motion of a hole is unfrustrated (frustrated) in an RVB (ferromagnetic) background.
The ground state of this model is rigorously proven to be a nearest-neighbor RVB state with a delocalized holon.
Such a graph has a property that the product of hopping matrix elements around any {\it cycle} (a loop of length $l\geq3$ in which only the first and the last vertices are equal) is {\it negative}. 
We also remark that the system is integrable thanks to the existence of extensive number of  conserved quantities  -- this is an example of 
Hilbert space fragmentation \cite{yang2020fragmentation, sala2020fragmentation, moudgalya2021fragmentation}.

\underline{A hole in a triangle}. We start by solving the two-electron problem for the Hubbard model on a triangle
with $U=\infty$  and $t>0$:
\begin{align}
    H = -t\sum_{i=1}^3\sum_{\sigma=\uparrow,\downarrow}\left [c^{\dagger}_{i,\sigma}c_{i+1,\sigma} + {\rm H.c.} \right] + \left [U=\infty \right ],
\end{align}
where the site $i=4$ is identified with $i=1$ ($c_{4,\sigma} \equiv c_{1,\sigma}$).
In the total $S= 1$ (triplet) sector, energy eigenvalues are $E_n = 2t \cos(\frac{2\pi n}{3})$, where $n=0,1,2,$ with three-fold degeneracies due to the spin-rotational symmetry (corresponding to the total $S^z = \pm 1,0$).
In the $S=0$ (singlet) sector, energy eigenvalues are $E_n = -2t \cos(\frac{2\pi n }{3})$, where $n=0,1,2.$
The ground state is the singlet state:
 \begin{align}
 \label{eq:single triangle}
     \left |{\rm GS} \right >
     = \frac 1 {\sqrt 3} \bigg ( \big | \VBone \big > +\left | \VBtwo \right >+\left | \VBthree \right > \bigg ),
 \end{align}
 where a circle on a vertex denotes the location of a hole and the magenta bond denotes the singlet state on two sites. 
The singlet state is oriented in a counter-clockwise direction on a triangle.
In the $S=0$ ground state, the hole's kinetic energy has its minimum possible value $-2t$, whereas it is frustrated in a spin-polarized background, with the lowest energy being $-t$.

Indeed, in the singlet subspace ($S^2=0$), unique basis states can be identified with the location of the holon, i.e. the state $\big | \VBone \big >$ can be identified as the state with a holon (with its creation operator $h_i^\dagger$) at the circled site. 
In the triplet sector ($S=1$), with a fixed total $S^z =\pm 1,0$, the basis states can similarly be identified by the position of the hole.
It is then easy to see that the  Hamiltonian of a hole in the singlet sector is given by $H_{\rm eff}^{(s)} = -t \sum_{i=1}^{3}\left (h^{\dagger}_{i}h_{i+1} +{\rm H.c.} \right ),$ whereas in the triplet sector with a fixed total $S^z$, $H_{\rm eff}^{(t)} = +t \sum_{i=1}^{3}\left (h^{\dagger}_{i}h_{i+1} +{\rm H.c.} \right ) = -t  \sum_{i=1}^{3}\left ({\rm e}^{-i\pi} h^{\dagger}_{i}h_{i+1} +{\rm H.c.} \right ).$
Effectively, the hole sees a $\pi$-flux through the triangle when the background spins form a triplet pair 
\footnote{This is true even when $t$ varies among different bonds, and when on-site chemical potential disorder and 
spin-independent interaction terms of the form Eq. \ref{eq:interactions} are present.}.

\begin{figure}[t]
    \centering
    \includegraphics[width = 0.45 \textwidth]{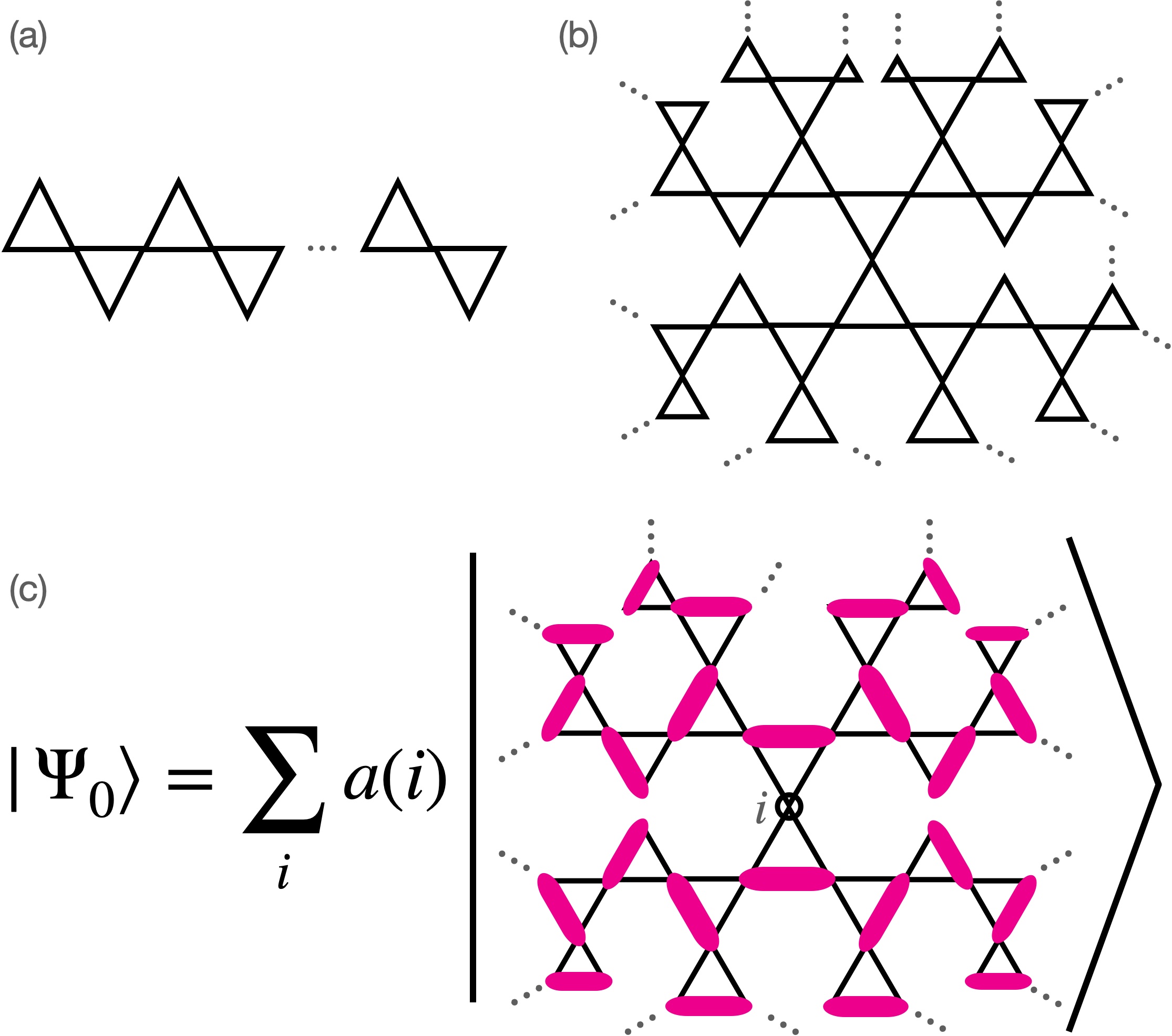}
    \caption{(a) Sawtooth geometry. (b) An example of a triangular cactus (or a  Husimi cactus). It is also possible that three or more triangles share the same vertex.
    (c) The RVB ground state induced by the hole motion in the $U=\infty $ Hubbard model on the triangular cactus. Here, $i$ denotes the location of the holon (circled) and magenta ellipses indicate singlet valence bonds of two $S=\frac 1 2$ spins.
    The amplitude, $a(i)$, of each valence-bond configuration is all positive, $a(i)>0$, with the counter-clockwise orientation of valence-bonds as introduced below Eq. \ref{eq:single triangle} \& \ref{eq:basis}. 
    }
    \label{fig:fig1}
\end{figure}

\underline{Triangular cactus}. We now consider the {\it single hole} problem in the $U=\infty$ Hubbard model on a triangular cactus. 
A triangular cactus is a planar graph where the only cycles -- loops of length $l\geq3$ in which only the first and the last vertices are equal --
are triangles and any edge belongs to a cycle.
Such a graph has previously been widely studied in the context of spin model (e.g. Heisenberg model) \cite{chandra1994Husimi, Tchernyshyov2009fermionic, Tchernyshyov2010structure}.
Fig. \ref{fig:fig1} (a-b) are 
examples of 
a triangular cactus.
We consider the following $U=\infty$ Hubbard models on such graphs with negative  but otherwise arbitrary hopping matrix elements $-t_{ij} < 0$:
\begin{align}
\label{eq:Hamiltonian}
    H =& -\sum_{\left <i,j \right >,\sigma} t_{ij}c^{\dagger}_{i,\sigma}c_{j,\sigma} \  +V(\{n_i\})
     + [U=\infty].
\end{align}
Here, $\left <i,j \right >$ denotes the directed bond from the site $i$ to $j$ of the graph, and $n_i = n_{i,\uparrow}+n_{i,\downarrow}$ is the number operator on site $i$. $t_{ij}\neq 0 $ only for those bonds $\left <i,j \right >$ connected by the triangular cactus.
Note that the number of sites $i$ of the graph is always odd ($2N_f+1$) and the number of {\it directed} bonds $\left < i,j \right >$ is $6N_f$, where $N_f$ is the number of plaquettes (or faces) $f$.
$V(\{n_i\})$ denotes arbitrary on-site disorder and interaction terms:
\begin{align}
\label{eq:interactions}
    V(\{n_i\}) = \sum_i \epsilon_i n_i + \sum_{i,j} V_{ij}n_i n_j + \cdots.
\end{align}
At half-filling (one electron per site), there is a $2^{2N_f+1}$
spin degeneracy.
The main result of the paper (the Theorem below) is that the motion of a single hole lifts such degeneracy and induces the RVB ground state.

Before going into technical details, we first define the convenient many-body basis of the problem. 
For this, we make a direct contact with quantum dimer models \cite{rokhsarKivelson1988QDM, moessnerSondhi2001, misguich2002QDM, verresen2022unifying}, and consider the states of  hard-core (nearest-neighbor) dimers on a triangular cactus graph, with a single monomer (that is, all sites but one are touched by a dimer). 
Once the location of the momomer is specified, it is easy to see that there is a {\it unique} dimer covering, which has exactly one dimer fully contained in every triangle (see Fig. \ref{fig:fig1} (c) for the illustration of such a configuration).
Now consider the Hamiltonian describing the hopping of a monomer: 
\begin{align}
    H_{\rm hop} = - t \sum_{\triangle} \bigg ( \big | \DimerOne \big >\big < \DimerTwo \big | &+\left | \DimerTwo \right >\left < \DimerThree \right |  \nonumber \\ &+\left | \DimerThree \right >\big < \DimerOne \big |  +{\rm H.c.} \bigg ),
\end{align}
where a circle on a vertex denotes the location of monomer.
The dimer is colored black to differentiate it from a singlet bond.
In any step in which the monomer hops to a nearest-neighbor site, one dimer is moved, but in such a way that it remains interior to the same triangle.
{\it Thus, we can label the dimers uniquely by a plaquette index $f$, and this index is preserved under the specified dynamics.}

Now let us consider the corresponding electron problem.
Given the location of the hole, $i$, and the corresponding unique dimer covering, let $\mathscr {\tilde S}_f$ and $\mathscr {\tilde S}_f^z$ be the total spin and spin component in the $z-$direction, respectively, of the two electrons touched by the dimer contained in the plaquette $f$.
The two spins form either a singlet or triplet state: $\mathscr {\tilde S}_f = 0,1.$
$\mathscr {\tilde S}_f$ and $\mathscr {\tilde S}_f^z$ constructed in this way form 
an extensive set of {\it local} conserved quantities: $[\mathscr {\tilde S}_f, H] = [\mathscr {\tilde S}_f^z, H]=0 $ for all $f$.
{\it Note that $\mathscr {\tilde S}_f$ and $\mathscr {\tilde S}_f^z$ are different from $S_f$ and $S_f^z$, the total spin and the spin in $z$-direction of the three sites in $f.$} 
Finally, we form the following orthonormal basis states:
\begin{align}
\label{eq:basis}
    \left |i,\{\mathscr {\tilde S}_f\},\{\mathscr {\tilde S}_f^z\} \right >,
\end{align}
where $i=1,2,...,2N_f+1$ and $f=1,2,...,N_f$.
{\it Again, we choose to orient valence-bonds in counter-clockwise direction around each triangle, $f,$ whenever $\mathscr {\tilde S}_f =0$.}
 (This introduces 
 a sign convention for 
 resonating-valence-bond-type wave-functions \cite{liang1988RVBWF}.)
Of these basis states, the state corresponding to the unique valence-bond covering with the holon at site $i$ will be denoted by
\begin{align}
\label{eq:VBC}
    \left | i, {\rm VBC} \right > \equiv \left |i,\{\mathscr {\tilde S}_f \equiv 0\},\{\mathscr {\tilde S}_f^z \equiv 0\} \right >.
\end{align}
Then, the following theorem is the main result of this paper.

{\it Theorem}: The ground state of the Hamiltonian Eq. \ref{eq:Hamiltonian} in the presence of a single hole ($2N_f$ electrons on $2N_f+1$ sites) is unique and is the positive ($a(i) >0$) superposition of all the possible valence-bond coverings $\left | i, {\rm VBC} \right >$. 
This is the nearest-neighbor ``resonating-valence-bond (RVB) state'' with a delocalized holon: 
\begin{align}
\label{eq:ground state}
    \left | \Psi_0 \right > = \sum_i a(i) \left |i, {\rm VBC} \right >. 
\end{align}
 (see Fig. \ref{fig:fig1} (c) for the illustration of this RVB state.)

The Theorem can be easily proven with the following well-known lemma (see e.g., Ref. \cite{liebLoss1993fluxes}). 

{\it Lemma} (diamagnetic inequality): 
Consider a single particle hopping problem under a magnetic field on a general 2-edge-connected planar graph in the presence of an arbitrary on-site potential term: 
\begin{align}
\label{eq:flux}
    T[\{\phi_f \}] + V_0 \equiv -\sum_{\left < i,j\right >} t_{ij} {\rm e}^{-i \theta_{ij}} \left|i \right >\left<j \right | + \sum_i \epsilon_i \left | i \right >\left < i \right |,  
\end{align}
where we assume $t_{ij}>0$  and $\theta_{ij}$ is an induced Berry phase on an edge $\left <i,j \right>$ due to a flux $\phi_f$ through a plaquette $f$ to which  $\left <i,j \right>$ belongs.
We will simply denote by $T_0$ the hopping matrix in the absence of a magnetic field: $T_0\equiv T[\{\phi_f \equiv 0\}].$
Here, a 2-edge-connected graph is a connected graph in which 
every edge belongs to at least one plaquette.
Formally, it is defined to be a connected graph that cannot be disconnected by deleting any single edge.
{\it Then, the flux configuration that minimizes the ground state energy of\  $T[\{\phi_f \}]$ is \underline{unique} and is the one without any flux: $\phi_f=0$ for all $f$, i.e. when $T[\{\phi_f\}] =T_0.$ } 
The physical meaning is that ``a magnetic field raises the energy."

{\it Proof of the lemma}: 
Let $\left | \psi' \right >$ be the normalized ground state of $T[\{\phi_f \}]+V_0$ for a given non-trivial flux configuration $\{\phi_f\}$ with the energy $E_0'$, and $\left | \psi \right >$ be the normalized ground state of $T_0 +V_0$ with the energy $E_0.$ 
It is easy to see that $E_0 \leq E_0'$ by using the triangle inequality:
\begin{align}
\label{eq:triangle ineq}
  E_0' 
&= -\sum_{\left < i,j\right >} t_{ij} {\rm e}^{-i\theta_{ij}} \psi_i'^{*} \psi'_j  +\sum_i \epsilon_i |\psi'_i|^2
 \nonumber \\
&\geq  -\sum_{\left < i,j\right >} t_{ij} |\psi_i'|\cdot |\psi_j'|+\sum_i \epsilon_i |\psi'_i|^2  \nonumber \\
&= \left < |\psi'| \right | \left (T_0 + V_0 \right ) \left | |\psi'| \right > \geq  E_0 .
\end{align}
Here, $|\cdot|$ denotes the matrix with every entry replaced by its absolute value: e.g., $(|A|)_{ij} \equiv |A_{ij}|$.

In order to prove the uniqueness, it is enough to show that the first inequality above is a strict inequality.
Let us assume otherwise, in which case each term in $-\left < \psi' \right | T[\{ \phi_f\}] \left | \psi' \right >$ is real and positive:
\begin{align}
\label{eq:positivity}
    {\rm e}^{-i\theta_{ij}} \psi_i'^* \psi'_j >0
\end{align}
for all $\left <i,j \right >$.
Now, let $\phi_f \neq 0$ for some plaquette $f,$ with its vertices $i_1,i_2,...,i_n$ ($i_{n+1 } \equiv i_1$).  From Eq. \ref{eq:positivity}, we obtain 
\begin{align}
    \prod_{k=1}^n {\rm e}^{-i\theta_{i_k i_{k+1}}} \psi_{i_k}'^* \psi'_{i_{k+1}} = {\rm e}^{-i \phi_f} \prod_{k=1}^n |\psi'_{i_k}|^2 >0,
\end{align}
which is in contradiction to the assumption that $\phi_f \neq 0$. 
This completes the proof. $\square$

{\it Proof of the Theorem}:
Since $[\mathscr {\tilde S}_f, H] = [\mathscr {\tilde S}_f^z, H]=0$, let us consider the Hamiltonian Eq. \ref{eq:Hamiltonian} in a given $\{\mathscr {\tilde S}_f\}$ and $\{\mathscr {\tilde S}_f^z\}$ sector,
$H \rvert_{\{\mathscr {\tilde S}_f\},\{\mathscr {\tilde S}_f^z\}}$. 
As shown in the single triangle problem above, the hole sees effective $\pi-$fluxes (no-fluxes) on triangles, $f$, at which $\mathscr {\tilde S}_f$ is a triplet (singlet).
{\it Hence, $H \rvert_{\{\mathscr {\tilde S}_f\},\{\mathscr {\tilde S}_f^z\}}$ is the Hamiltonian of a single hole hopping problem in the presence of $\pi-$fluxes through the triangle plaquettes, $f,$ with $\mathscr {\tilde S}_f =1.$}
According to the lemma (diamagnetic inequality), the energy minimizing flux configuration is unique and is the one without any flux, and hence, $\mathscr {\tilde S}_f=0$ and $\mathscr {\tilde S}_f^z=0$ for all $f$.
Also, 
\begin{align}
\label{eq:H_restricted}
    H\rvert_{\{\mathscr {\tilde S}_f=0\},\{\mathscr {\tilde S}_f^z=0\}} = -\sum_{\left <i,j \right > } t_{ij} \left | i \right > \left < j \right | + \sum_{i} \tilde V_i \left | i \right > \left < i \right |,
\end{align}  
where $\tilde V_i \equiv V(\{ n_i =0, n_{j \neq i} = 1\})$ is the effective on-site potential felt by the hole at site $i$. 
Since the off-diagonal elements of $H\rvert_{\{\mathscr {\tilde S}_f=0\},\{\mathscr {\tilde S}_f^z=0\}}$ are all negative, 
the Perron-Frobenius theorem ensures that the ground state, $\left | \Psi_0 \right >,$ of $H\rvert_{\{\mathscr {\tilde S}_f=0\},\{\mathscr {\tilde S}_f^z=0\}}$ (and hence of $H$) is the superposition of all the basis states (Eq. \ref{eq:VBC}) with  positive coefficients, Eq. \ref{eq:ground state}.
$\square$

{\it $t-J$ model}.
The nearest-neighbor RVB state of the form Eq. \ref{eq:ground state} with $a(i)>0$ is still a ground state in the  presence of nearest-neighbor antiferromagnetic Heisenberg interactions, $J >0$ of the following form:
\begin{align}
\label{eq:Heisenberg}
    H_J &= \sum_{f} J_{f} \sum_{l=1}^3 \vec S_l^{(f)} \cdot \vec S_{l+1}^{(f)}  \nonumber \\
    &=\sum_f  \frac {J_f} 2 \left [ S_f (S_f +1) - \frac 3 4 n_f \right ].
\end{align}
Here, $\vec S^{(f)}_l =\sum_{s,s'=\uparrow,\downarrow}c^{\dagger}_{l,s} \frac{\vec \sigma_{ss'}}{2} c_{l,s'}$ ($l=1,2,3$) is the spin operator on site $l$ of a triangle $f$ (with $S^{(f)}_4\equiv S^{(f)}_1$),  $\vec S_f = \sum_{l=1 }^3 \vec S_l^{(f)}$, and $n_f$ is the total number operator on a triangle $f$. Antiferromagnetic interactions $J$ are uniform for bonds of the same triangle $f$, while they can differ on different triangles.

{\it Proof of the Theorem in the presence of $J\geq 0$}: 
Observe that each $\left | i , {\rm VBC} \right >$ describing a valence-bond covering with the holon at site $i$ is an eigenstate of $H_J$ with the lowest possible energy eigenvalue (for a fixed $i$): 
\begin{align}
    H_J \left | i , {\rm VBC} \right >  
    = \left (-\frac 3 4\sum_{f }J_f \right )\left | i , {\rm VBC} \right >.
\end{align}
This means that the ground state of the total Hamiltonian including $H_J$ is still in the $\{ \mathscr {\tilde S}_f =0\}$ sector.
Moreover, since $H_J \rvert_{\{ \mathscr {\tilde S}_f =0\},\{ \mathscr {\tilde S}_f^z =0\} }$ is diagonal in the basis $\left | i , {\rm VBC } \right >,$ it follows from the Perron-Frobenius theorem that the ground state is still of the form Eq. \ref{eq:ground state}, with $a(i)$ modified but remaining positive. $\square$

{\it Integrability}.
When $J = 0$ (i.e., $H_J = 0$), the entire excited state spectra of Eq. \ref{eq:Hamiltonian} can be obtained by exploiting the extensive set of quantum numbers $\{\mathscr {\tilde S}_f\}$ and $\{\mathscr {\tilde S}_f^z\}$ ($f = 1,2,...,N_f$). 
The spin excitations are $\mathscr {\tilde S}_f=1$ triplets localized on certain triangles $f$.
Let us denote by $\triangle_s$ ($\triangle_t$) the set of directed bonds of triangles at which $\mathscr {\tilde S}_f$ forms a singlet (triplet). 
The charge spectrum can be obtained by diagonalizing the single hole problem in the presence of $\pi-$fluxes on $\triangle_t$ \footnote{Similar reasoning is used in obtaining anyon states in the Kitaev model on the honeycomb lattice \cite{kitaev2006anyons}}:
\begin{align}
    H\rvert_{\{\mathscr {\tilde S}_f\},\{\mathscr {\tilde S}_f^z\}} = -\sum_{\left <i,j \right > \in \triangle_s } t_{ij} \left | i \right > \left < j \right |  &-\sum_{\left <i,j \right > \in \triangle_t } t_{ij} {\rm e}^{-i\pi} \left | i \right > \left < j \right | \nonumber \\
    & + \sum_{i} \tilde V_i \left | i \right > \left < i \right |.
\end{align}
In the presence of $H_J,$ $\mathscr {\tilde S}_f$ and $\mathscr {\tilde S}_f^z$ are no longer good quantum numbers, and the system is no longer integrable.

{\it Relevance of the sign of hopping matrix elements}. In the presence of the uniform $\pi-$flux on each triangle, which amounts to changing the sign of hopping terms $t_{ij} \rightarrow -t_{ij}$, the ground state manifold consists of the states with $N_f$ uncorrelated spin triplets, each of which is localized on the triangle $f$:
\begin{align}
\label{eq:triplet GS}
    \left | \{\mathscr {\tilde S}_f^z\} \right > \equiv \sum_i b(i) \left |i,\{\mathscr {\tilde S}_f=1\},\{\mathscr {\tilde S}_f^z\} \right >,
\end{align}
where $b(i)>0$ and $\{\mathscr {\tilde S}_f^z\} =\pm 1,0.$ The ground states are $3^{N_f}$-fold degenerate;
among them is the familiar fully-polarized Nagaoka ferromagnet. 
If the $\pi-$fluxes are present only in some $N_{\phi}(<N_f)$ number of triangles, the ground state manifold consists of the states with localized triplets $\mathscr {\tilde S}_f =1$ on those $N_{\phi}$ triangles and is $3^{N_{\phi}}$-fold degenerate.

{\it Spin-$\frac{1}{2}$ bosons}. All of the above conclusions remain true for spin-$\frac 1 2$ hard core bosons if the sign of the hopping term is reversed.
This is a weak converse to the results of Ref. \cite{eisenberg2002bosonFerromagnetism, yang2003bosonFerromagnetism} which show that the ground state of spin$-\frac 1 2$ bosons is a fully-polarized ferromagnet when the hopping matrix elements are all negative.

\underline{Discussion}.
The exact solvability of the present model relies on its ``tree-like'' structure, i.e. due to the absence of loops other than triangles.
Exact generalization of this result to a 2D or higher dimensional lattice is likely to be obstructed by the existence of longer-ranged valence-bonds generated by the hopping of a holon around an additional loop adjacent to a certain triangle.
Moreover, the existence of additional even-length loops produces a tendency towards a ferromagnetism, as exemplified by the Nagaoka's theorem on a bipartite lattice, and  frustrates a tendency to a singlet formation, making analytic solution highly unlikely.    
However, if the number of non-triangular loops is suppressed in comparison to the number of (corner-sharing) triangles, it is likely that a version a short-ranged RVB state  is stabilized: a kagome lattice or a suitably decorated version of it may be such an example.
Such an idea is in line with the attempts to reproduce quantum dimer models as a limiting case by suitably decorating each edge of $2$D lattices with Majumdar-Ghosh chain \cite{raman20052, moessner2006decoration}.


We hope that the present exact result will prove to be a fruitful starting point for a numerical search for a doping-induced RVB state (as opposed to doping an RVB state induced by frustrated antiferromagnetic interactions).
In particular, a numerical study of the $U=\infty$ Hubbard model on a kagome lattice is currently lacking,
although such studies have been carried out for the square and triangular lattices \cite{liu2012phases, zhu2022doped}.
Whether doping dilute holes in the $U=\infty$ Hubbard model on the kagome lattice leads to superconductivity
\cite{Senthil_FL*, sachdev2016FL*, senthil2000SC*, senthil2001SC*2}, a holon Fermi liquid, a holon Wigner crystal \cite{jiang2017holonWC}, or some other state 
 is an interesting open question.






\section{acknowledgments}
I deeply appreciate Steve Kivelson for his support and generosity during this project and also for providing extensive comments and suggestions on the draft.
I'm indebted to Nicholas O'Dea and Andrew Yuan for helpful discussions, and thank Zhaoyu Han, Chaitanya Murthy, Hong-Chen Jiang, Akshat Pandey and John Sous for collaboration on related topics. 
I appreciate Ashvin Vishwanath, 
Dung-Hai Lee,   Roderich Moessner, Elliot Lieb, Sriram Shastry, Josephine Yu, and especially Ruben Verresen and Ronny Thomale for helpful comments on the draft.
This work was supported, in
part, by NSF grant No. DMR-2000987 at Stanford University.

\bibliography{ref.bib}

\end{document}